\documentclass[apm,reprint,amsmath,amssymb]{revtex4-1}

\usepackage{graphicx}
\usepackage{bm}

\begin{document}

\title{Electronic structure and the origin of the high ordering temperature
in SrRu$_2$O$_6$}

\author{David J. Singh}

\affiliation{
Materials Science and Technology Division,
Oak Ridge National Laboratory, Oak Ridge, Tennessee 37831-6056}

\date{\today}

\begin{abstract}
SrRu$_2$O$_6$ is a layered honeycomb lattice material with an
extraordinarily high magnetic ordering temperature.
We investigated this material using density functional calculations.
We find that the energy scales for moment formation and ordering
are similar and high. Additionally, we find that the magnetic
anisotropy is high and favors moments oriented along the $c$-axis.
This provides an explanation for the exceptionally high ordering
temperature.
Finally, the compound is found to be semiconducting at the bare density
functional level, even without magnetic order.
Experimental consequences of this scenario for the high ordering
temperature are discussed.
\end{abstract}

\pacs{}

\maketitle

\section{Introduction}

Hiley and co-workers recently reported synthesis of the layered honeycomb
lattice oxide SrRu$_2$O$_6$, which contains pentavalent Ru$^{5+}$ ions in
octahedral coordination. \cite{hiley}
The compound has antiferromagnetic ordering with an ordering temperature
above 500 K, which is an extremely high value, particularly considering the
layered crystal structure. In fact, while a number of remarkably
high magnetic ordering temperature 4d and 5d oxides have been
discovered, most notably SrTcO$_3$, CaTcO$_3$ and NaOsO$_3$,
\cite{rodriguez,avdeev,shi,calder}
SrRu$_2$O$_6$ is the first example of an apparently 2D material
in this category, and in fact its ordering temperature exceeds
that of NaOsO$_3$.

\begin{figure}
 \includegraphics[width=\columnwidth,angle=0]{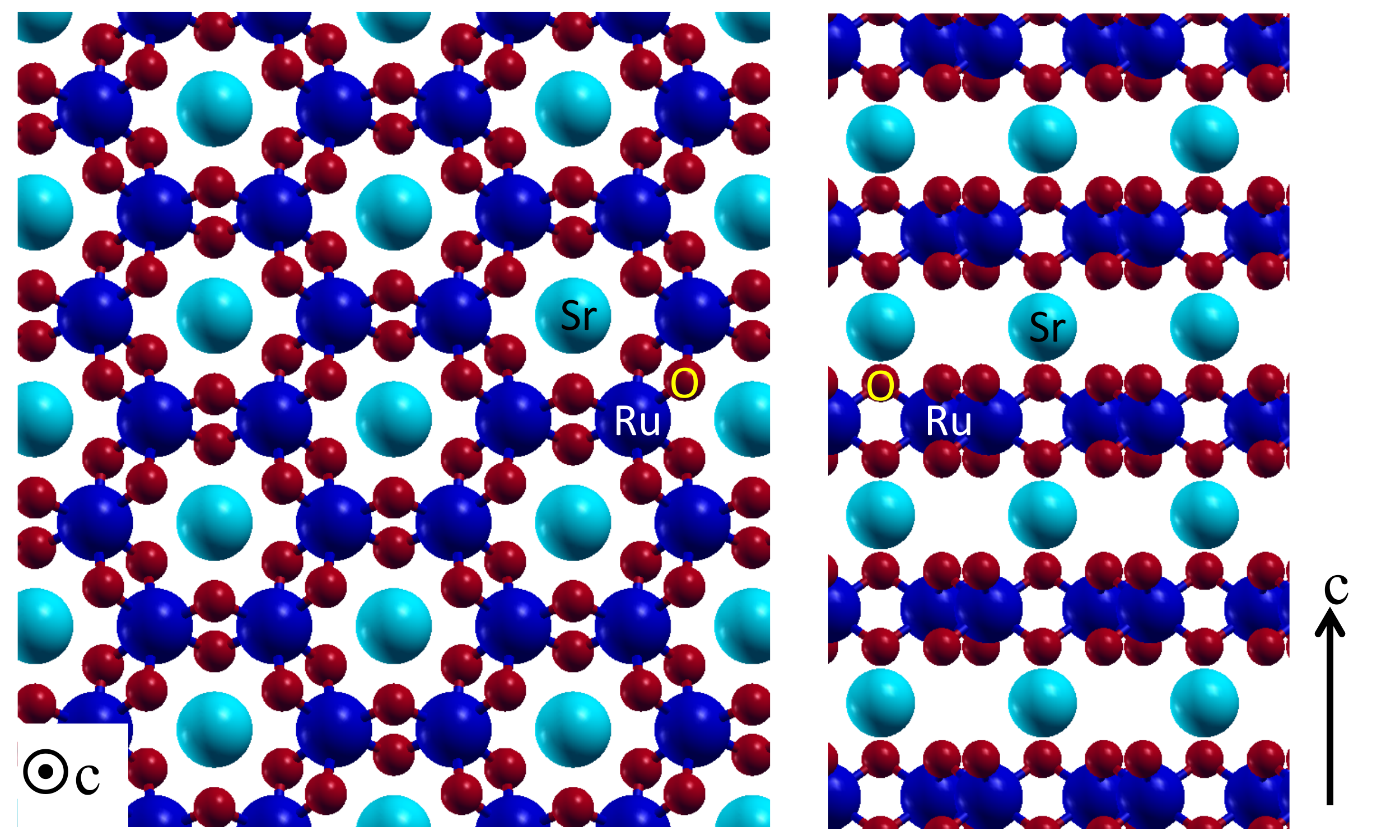}
\caption{Crystal structure of SrRu$_2$O$_6$ showing the honeycomb
lattice planes separated by Sr ions. }
\label{structure}
\end{figure}

The crystal structure of SrRu$_2$O$_6$ consists of
honeycomb lattice planes
of Ru$^{5+}$ ions, stacked directly on top of each other with
intervening Sr$^{2+}$ to form
a hexagonal lattice, as shown in Fig. \ref{structure}.
There is one formula unit (two Ru atoms) per unit cell.

\section{Approach}

We did density functional calculations using the experimental
crystal structure, which was determined by synchrotron
xray and neutron diffraction. \cite{hiley} The accuracy
of this structure is supported
by the fact that our calculated forces in the antiferromagnetic
ground state with this structure are
below 4 mRy/bohr. This is essentially zero at the precision of density
functional calculations. The calculations were done using the
general potential linearized augmented planewave (LAPW) method
\cite{singh-book} as implemented in the WIEN2k code. \cite{wien2k}
We used LAPW sphere radii of 2.05 bohr for Sr and Ru and 1.55 bohr for
O. We used well converged LAPW basis sets and included local orbitals
\cite{singh-lo}
for the semi-core states of Sr and Ru.

We did calculations both in a scalar
relativistic approximation and with inclusion of spin-orbit,
and find similar results.
The calculations were done using the generalized gradient approximation
(GGA) of Perdew, Burke and Ernzerhof (PBE). \cite{pbe}

\begin{figure}
 \includegraphics[width=\columnwidth,angle=0]{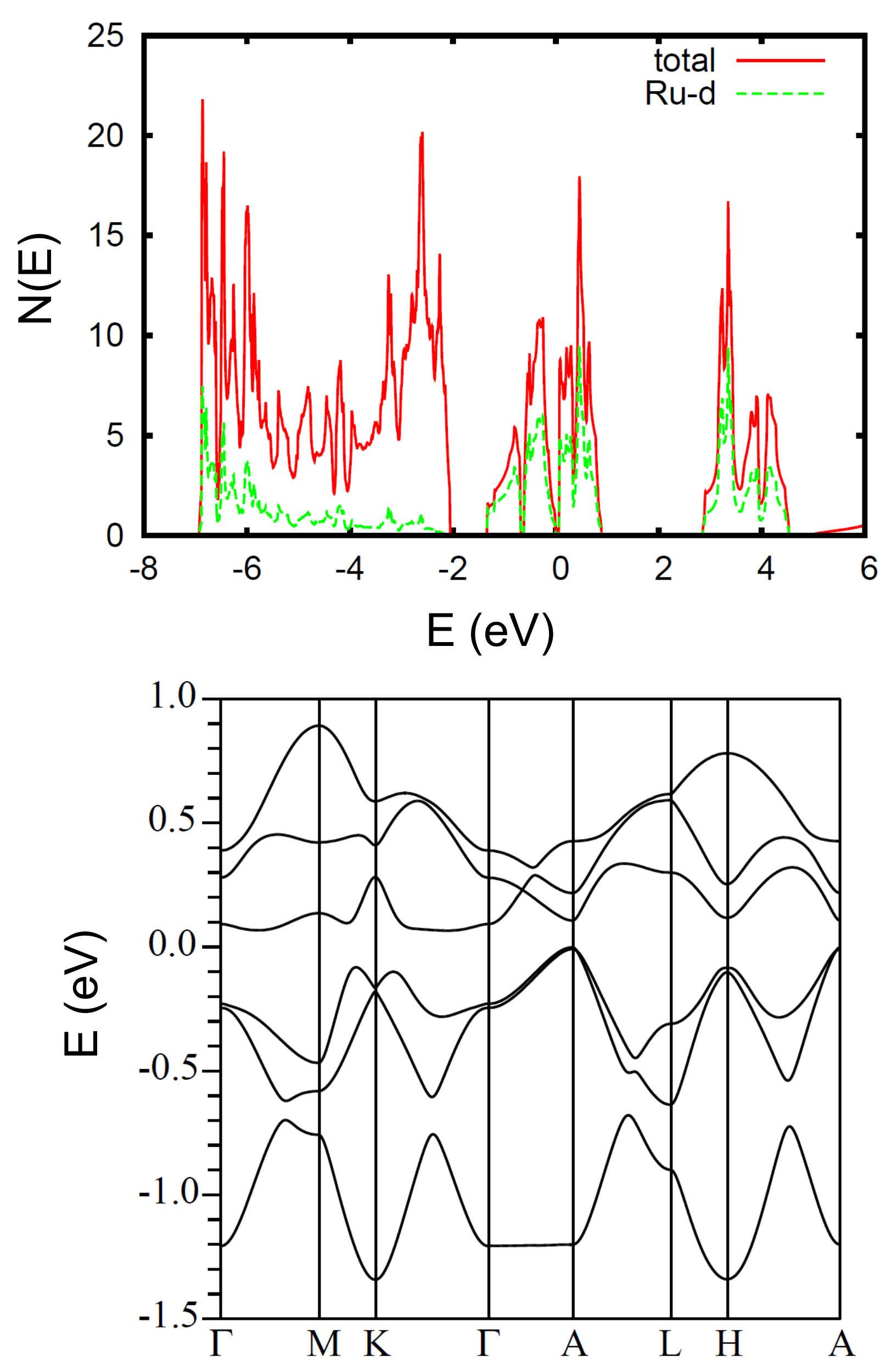}
\caption{Electronic density of states and Ru $d$ projection (top)
and band structure in the range around the Fermi level (bottom) as obtained
in non-spin-polarized PBE GGA calculations, including spin orbit.
The valence band maximum is set to 0 eV.}
\label{dos}
\end{figure}

\section{Results and Discussion}

We start with the electronic structure. The calculated density of states
without spin polarization as obtained with the PBE GGA is shown in 
Fig. \ref{dos}, along with the band structure near the Fermi level.
As expected, the electronic structure shows Ru$^{5+}$, with a half-filled
Ru $t_{2g}$ derived manifold. Since there are two Ru ions per unit cell,
there are six $t_{2g}$ bands and a band gap is possible without magnetism
even though there are an odd number of $t_{2g}$ electrons per atom.
This is the case.
The calculated non-spin-polarized band gap is 0.06 eV,
including spin-orbit and 0.05 eV in a scalar relativistic approximation.
Importantly, substantial hybridization between Ru 4$d$ and O $2p$ is
evident in the Ru $d$ projected density of states. For example, there
is substantial Ru $d$ character in the O 2$p$ bands, especially at
the bottom, but extending almost to the top of this manifold.

The honeycomb lattice is not frustrated against near neighbor
antiferromagnetism.
We did spin-polarized calculations for various ordering patterns.
These were the near neighbor antiferromagnetic state, in which neighboring Ru
in plane are antiferromagnetically aligned, and the $c$-axis stacking
is also antiferromagnetic (denoted AF1), the same in-plane order, but
stacked ferromagnetically along the $c$-axis (denoted AF2), a ferromagnetic
order (denoted F), and ferromagnetic planes stacked antiferromagnetically
(denoted AF3).

\begin{figure}
 \includegraphics[width=\columnwidth,angle=0]{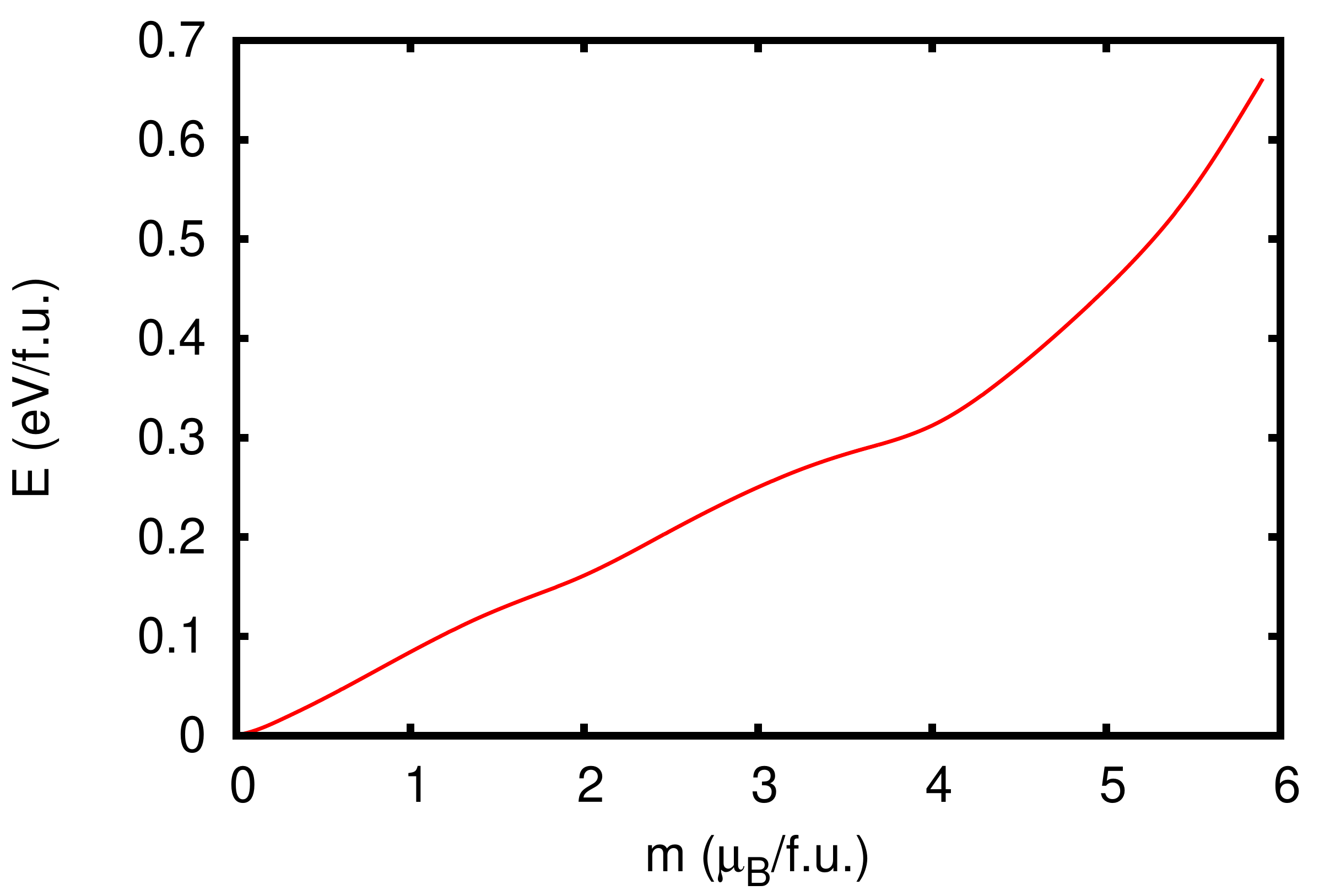}
\caption{Fixed spin moment energy as a function of spin magnetization
on a per formula unit basis as obtained with the PBE GGA.}
\label{fsm}
\end{figure}

Neither of the orders with ferromagnetic planes (F or AF3) yielded a 
spin-polarized solution with the PBE GGA.
This was confirmed by fixed spin moment calculations (Fig. \ref{fsm}).
These show a monotonically increasing energy with constrained 
spin magnetization. The fixed spin moment curve shows a roughly
linear increase in energy with magnetization at low magnetizations,
reflecting the presence of a band gap.
We note that the strong hybridization with O is evident in the fixed
spin moment results. In particular, with an imposed ferromagnetic
spin magnetization of 3 $\mu_B$/Ru only $\sim$ 1.8 $\mu_B$ is
in the Ru LAPW sphere (radius 2.05 bohr). Considering the
extent of the Ru 4$d$ atomic orbitals, the implication is that
roughly 1 $\mu_B$/Ru, i.e. 1/3 of the total imposed magnetization
lies on the O atoms. This is qualitatively similar to the Ru$^{5+}$
double perovskite oxide, Sr$_2$YRuO$_6$, \cite{mazin-ru}
and the Ru$^{4+}$ ferromagnet, SrRuO$_3$. \cite{singh-ru}

\begin{figure}
 \includegraphics[width=\columnwidth,angle=0]{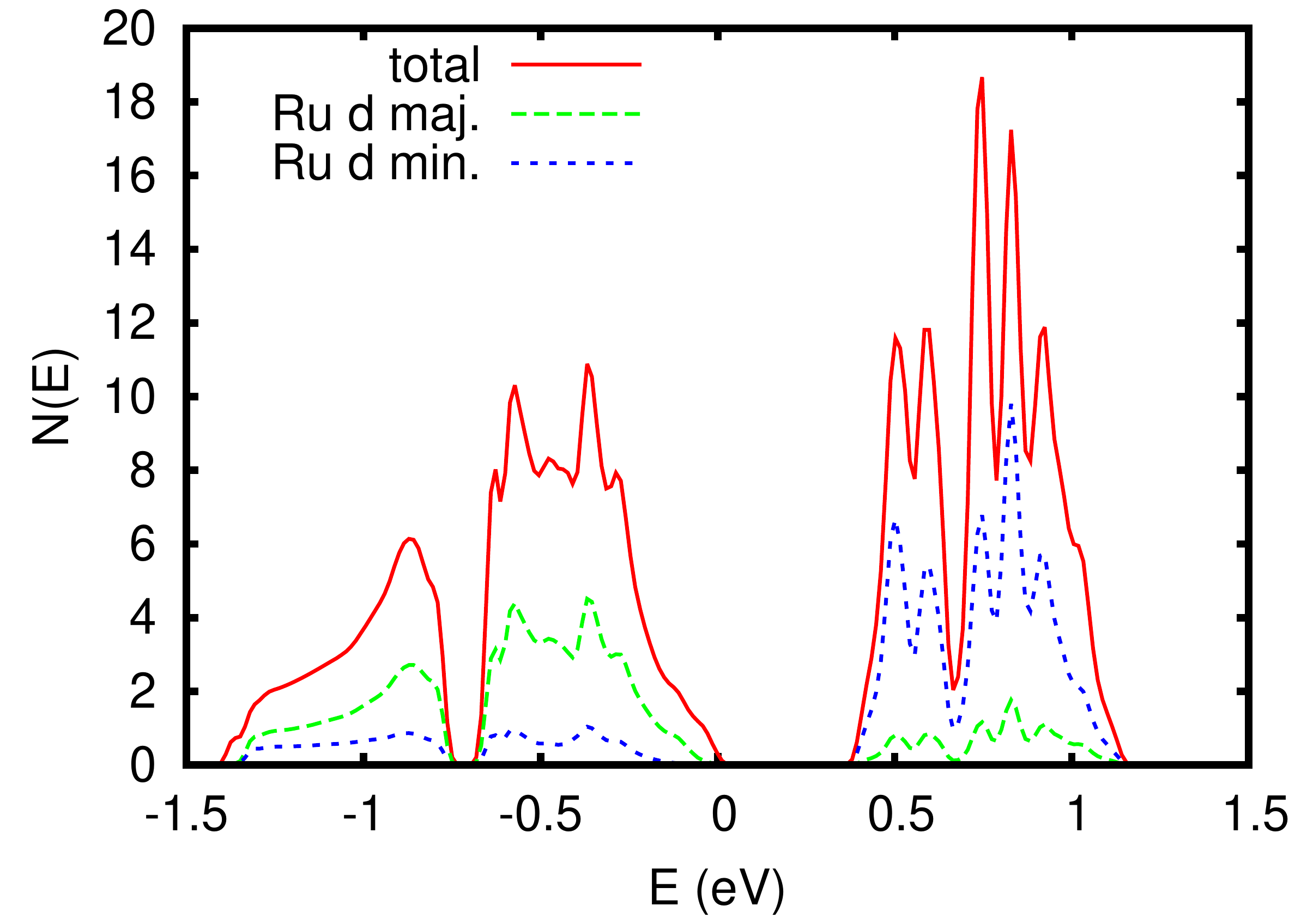}
\caption{Density of states for the AF1 ground state including spin orbit.}
\label{dos-afso}
\end{figure}

On the other hand,
we find very stable AF1 and AF2 orderings. The lowest energy AF1 order,
is 0.20 eV lower in energy per formula unit than the non-spin-polarized
case, while the AF2 order is only 0.003 eV per formula
unit higher than the ground state.
The small energy difference between the AF1 and AF2 states
means that the out-of-plane interactions are very weak compared to the
in-plane ordering energy. Low dimensional magnets, as defined
in terms of low interlayer couplings relative to in-plane
couplings, can have suppressed
ordering temperatures, usually logarithmically in the ratio of the
out-of-plane to in-plane magnetic interactions. \cite{kt}
This is expected to be the case for Heisenberg or XY moments, but
not for Ising like moments. We find that SrRu$_2$O$_6$ has a strong
magnetic anisotropy that favors moment directions along the $c$-axis.
For the AF1 ground state, we find that the energy with moments along the 
$c$-axis is 2.8 meV per formula unit lower than with moments oriented
along the $a$-axis in PBE-GGA calculations with spin orbit.
Therefore, a suppression of the ordering due to the layered structure
is not expected even though the interlayer magnetic interactions are weak.

The calculated spin
moment in the Ru sphere for the AF1 ground state is 1.3 $\mu_B$/Ru,
even lower than the induced moment in the fixed spin moment calculations.
Nonetheless the ordering opens a sizable gap in the $t_{2g}$ bands.
The band gap for the AF1 ordering with the PBE GGA is 0.43 eV without
spin orbit. With spin orbit, there is an orbital moment,
opposite to the spin moment following Hund's rule, of size
0.09 $\mu_B$ and the PBE-GGA band gap is 0.39 eV.
The $t_{2g}$ density of states is shown in Fig. \ref{dos-afso}.
The moment on the Ru of 1.3 $\mu_B$
is strongly reduced relative to the nominally
expected spin moment of 3 $\mu_B$ for a half-filled $t_{2g}$ band. Based on
the fixed spin moment results and the strong covalency we infer that most of
this reduction is a consequence of covalency between the Ru 4$d$ and O 2$p$
states. This is similar to recent results for the 5$d$
double perovskite Sr$_2$ScOsO$_6$. \cite{taylor}
We note that this is a mechanism that has been discussed previously,
\cite{jacobson,tofield}
but appears to be
particularly large for these more covalent $4d$ and $5d$ materials.
 
This covalency between Ru and O is important for understanding the high
energy scale associated with magnetic ordering, which in turn provides
an explanation for the high ordering temperature.
Magnetism is much more common in 3$d$
oxides than in 4$d$ and 5$d$ oxides.
Because of this it is often presumed that the magnetism of 4$d$ and
5$d$ oxides is inherently weak. However, this is clearly not the case,
as is evident when one considers the ferromagnetism
of metallic SrRuO$_3$ \cite{longo,bouchard}
and the
very high ordering temperature in SrTcO$_3$. \cite{rodriguez}
Actually, as is well known, magnetism arises from intersite coupling
of moments on ions. In oxides, as in other materials, strong intersite
coupling of moments is favored by strong covalency.
\cite{anderson,goodenough}

Most magnetic materials are described within a local moment picture,
in which moments that exist due to on-site atomic interactions independent
of ordering are subject to inter-site couplings that determine the ordering
temperature. The reason why most 4$d$ and 5$d$ oxides are not magnetic
is not that these interactions, which would determine the ordering
temperature are weak.
Rather it is because that these elements have more extended $d$
orbitals than 3$d$ transition metals.
This leads to lower onsite Coulomb integrals and
stronger covalency, both of which work against formation of local moments.
Thus as covalency is increased, one expects the intersite interactions,
and the ordering temperature to increase so long as moments can form,
and then to vanish with the disappearance of the moments. In the region
of highest ordering temperature the energy scales for moment formation and
for ordering the moments will be comparable and therefore the existence of
moments will depend on the ordering. For metallic
magnets this is the itinerant limit. \cite{rhodes}

The elemental ferromagnets, Fe, Co and
Ni have Curie temperatures of 1043 K, 1400 K and 627 K, respectively.
Taking into account the different moments of 2.1 $\mu_B$, 1.6 $\mu_B$, 
and  0.6 $\mu_B$, respectively, one observes that the relative
ordering strength increases strongly
as the system becomes more itinerant, i.e. going from Fe to Co to Ni.
\cite{gunnarsson}
Thus it can be seen that increasing itinerancy favors increasing
Curie temperature.
The same principle is operative here.
In fact this has been discussed previously in the context of
SrRuO$_3$,
CaTcO$_3$ and SrTcO$_3$ based on density functional calculations,
\cite{mazin-ru,rodriguez,avdeev} and
subsequently for SrTcO$_3$,
in terms of dynamical mean field calculations with similar
conclusions. \cite{mravlje}

\section{Summary and Conclusions}

The extremely high ordering temperature of SrRu$_2$O$_6$ in a layered oxide
provides a new model system for exploring the interplay of covalency and
moment formation in a 4$d$ oxide.
The results suggest some experimental expectations
that may be tested. First of all, the comparable energy
scales for moment formation and ordering imply that the moments should
strongly
decrease as the ordering temperature is approached from below.
Secondly, the band gap should show a rather strong temperature dependence
near the ordering temperature, falling to a reduced value above the
ordering. These two expectations are similar to what is seen in 
NaOsO$_3$, \cite{calder}
except that in the present case, the non-magnetic case
is a small band gap semiconductor instead of a metal. In this sense
SrRu$_2$O$_6$ may provide an interesting exception to one of the standard
experimental
characterizations of a Mott insulating oxide, specifically an oxide with
an odd number of electrons per transition metal atom that
has an antiferromagnetic insulating ground state and stays insulating
above the magnetic ordering temperature.
Third, the reduction in the moments near the ordering temperature may
lead to unusual lattice behavior, such as an invar effect or even a
contraction as the ordering temperature is approached from below.

\acknowledgments

This work was supported by the Department of Energy, Basic Energy Sciences,
Materials Sciences and Engineering Division.

\bibliography{SrRu2O6}

\begin{thebibliography}{22}%
\makeatletter
\providecommand \@ifxundefined [1]{%
 \@ifx{#1\undefined}
}%
\providecommand \@ifnum [1]{%
 \ifnum #1\expandafter \@firstoftwo
 \else \expandafter \@secondoftwo
 \fi
}%
\providecommand \@ifx [1]{%
 \ifx #1\expandafter \@firstoftwo
 \else \expandafter \@secondoftwo
 \fi
}%
\providecommand \natexlab [1]{#1}%
\providecommand \enquote  [1]{``#1''}%
\providecommand \bibnamefont  [1]{#1}%
\providecommand \bibfnamefont [1]{#1}%
\providecommand \citenamefont [1]{#1}%
\providecommand \href@noop [0]{\@secondoftwo}%
\providecommand \href [0]{\begingroup \@sanitize@url \@href}%
\providecommand \@href[1]{\@@startlink{#1}\@@href}%
\providecommand \@@href[1]{\endgroup#1\@@endlink}%
\providecommand \@sanitize@url [0]{\catcode `\\12\catcode `\$12\catcode
  `\&12\catcode `\#12\catcode `\^12\catcode `\_12\catcode `\%12\relax}%
\providecommand \@@startlink[1]{}%
\providecommand \@@endlink[0]{}%
\providecommand \url  [0]{\begingroup\@sanitize@url \@url }%
\providecommand \@url [1]{\endgroup\@href {#1}{\urlprefix }}%
\providecommand \urlprefix  [0]{URL }%
\providecommand \Eprint [0]{\href }%
\providecommand \doibase [0]{http://dx.doi.org/}%
\providecommand \selectlanguage [0]{\@gobble}%
\providecommand \bibinfo  [0]{\@secondoftwo}%
\providecommand \bibfield  [0]{\@secondoftwo}%
\providecommand \translation [1]{[#1]}%
\providecommand \BibitemOpen [0]{}%
\providecommand \bibitemStop [0]{}%
\providecommand \bibitemNoStop [0]{.\EOS\space}%
\providecommand \EOS [0]{\spacefactor3000\relax}%
\providecommand \BibitemShut  [1]{\csname bibitem#1\endcsname}%
\let\auto@bib@innerbib\@empty
%</preamble>
\bibitem [{\citenamefont {Hiley}\ \emph {et~al.}(2014)\citenamefont {Hiley},
  \citenamefont {Lees}, \citenamefont {Fisher}, \citenamefont {Thompsett},
  \citenamefont {Agrestini}, \citenamefont {Smith},\ and\ \citenamefont
  {Walton}}]{hiley}%
  \BibitemOpen
  \bibfield  {author} {\bibinfo {author} {\bibfnamefont {C.~I.}\ \bibnamefont
  {Hiley}}, \bibinfo {author} {\bibfnamefont {M.~R.}\ \bibnamefont {Lees}},
  \bibinfo {author} {\bibfnamefont {J.~M.}\ \bibnamefont {Fisher}}, \bibinfo
  {author} {\bibfnamefont {D.}~\bibnamefont {Thompsett}}, \bibinfo {author}
  {\bibfnamefont {S.}~\bibnamefont {Agrestini}}, \bibinfo {author}
  {\bibfnamefont {R.~I.}\ \bibnamefont {Smith}}, \ and\ \bibinfo {author}
  {\bibfnamefont {R.~I.}\ \bibnamefont {Walton}},\ }\href@noop {} {\bibfield
  {journal} {\bibinfo  {journal} {Angew. Chem. Int. Ed.}\ }\textbf {\bibinfo
  {volume} {53}},\ \bibinfo {pages} {4423} (\bibinfo {year}
  {2014})}\BibitemShut {NoStop}%
\bibitem [{\citenamefont {Rodriguez}\ \emph {et~al.}(2011)\citenamefont
  {Rodriguez}, \citenamefont {Poineau}, \citenamefont {Llobet}, \citenamefont
  {Kennedy}, \citenamefont {Avdeev}, \citenamefont {Thorogood}, \citenamefont
  {Carter}, \citenamefont {Seshadri}, \citenamefont {Singh},\ and\
  \citenamefont {Cheetham}}]{rodriguez}%
  \BibitemOpen
  \bibfield  {author} {\bibinfo {author} {\bibfnamefont {E.~E.}\ \bibnamefont
  {Rodriguez}}, \bibinfo {author} {\bibfnamefont {F.}~\bibnamefont {Poineau}},
  \bibinfo {author} {\bibfnamefont {A.}~\bibnamefont {Llobet}}, \bibinfo
  {author} {\bibfnamefont {B.~J.}\ \bibnamefont {Kennedy}}, \bibinfo {author}
  {\bibfnamefont {M.}~\bibnamefont {Avdeev}}, \bibinfo {author} {\bibfnamefont
  {G.~J.}\ \bibnamefont {Thorogood}}, \bibinfo {author} {\bibfnamefont {M.~L.}\
  \bibnamefont {Carter}}, \bibinfo {author} {\bibfnamefont {R.}~\bibnamefont
  {Seshadri}}, \bibinfo {author} {\bibfnamefont {D.~J.}\ \bibnamefont {Singh}},
  \ and\ \bibinfo {author} {\bibfnamefont {A.~K.}\ \bibnamefont {Cheetham}},\
  }\href@noop {} {\bibfield  {journal} {\bibinfo  {journal} {Phys. Rev. Lett.}\
  }\textbf {\bibinfo {volume} {106}},\ \bibinfo {pages} {067201} (\bibinfo
  {year} {2011})}\BibitemShut {NoStop}%
\bibitem [{\citenamefont {Avdeev}\ \emph {et~al.}(2011)\citenamefont {Avdeev},
  \citenamefont {Thorogood}, \citenamefont {Carter}, \citenamefont {Kennedy},
  \citenamefont {Ting}, \citenamefont {Singh},\ and\ \citenamefont
  {Wallwork}}]{avdeev}%
  \BibitemOpen
  \bibfield  {author} {\bibinfo {author} {\bibfnamefont {M.}~\bibnamefont
  {Avdeev}}, \bibinfo {author} {\bibfnamefont {G.~J.}\ \bibnamefont
  {Thorogood}}, \bibinfo {author} {\bibfnamefont {M.~L.}\ \bibnamefont
  {Carter}}, \bibinfo {author} {\bibfnamefont {B.~J.}\ \bibnamefont {Kennedy}},
  \bibinfo {author} {\bibfnamefont {J.}~\bibnamefont {Ting}}, \bibinfo {author}
  {\bibfnamefont {D.~J.}\ \bibnamefont {Singh}}, \ and\ \bibinfo {author}
  {\bibfnamefont {K.~S.}\ \bibnamefont {Wallwork}},\ }\href@noop {} {\bibfield
  {journal} {\bibinfo  {journal} {J. Am. Chem. Soc.}\ }\textbf {\bibinfo
  {volume} {133}},\ \bibinfo {pages} {1654} (\bibinfo {year}
  {2011})}\BibitemShut {NoStop}%
\bibitem [{\citenamefont {Shi}\ \emph {et~al.}(2009)\citenamefont {Shi},
  \citenamefont {Guo}, \citenamefont {Yu}, \citenamefont {Arai}, \citenamefont
  {Belik}, \citenamefont {Sato}, \citenamefont {Yamaura}, \citenamefont
  {Takayama-Muromachi}, \citenamefont {Tian}, \citenamefont {Yang},
  \citenamefont {Li}, \citenamefont {Varga}, \citenamefont {Mitchell},\ and\
  \citenamefont {Okamoto}}]{shi}%
  \BibitemOpen
  \bibfield  {author} {\bibinfo {author} {\bibfnamefont {Y.~G.}\ \bibnamefont
  {Shi}}, \bibinfo {author} {\bibfnamefont {Y.~F.}\ \bibnamefont {Guo}},
  \bibinfo {author} {\bibfnamefont {S.}~\bibnamefont {Yu}}, \bibinfo {author}
  {\bibfnamefont {M.}~\bibnamefont {Arai}}, \bibinfo {author} {\bibfnamefont
  {A.~A.}\ \bibnamefont {Belik}}, \bibinfo {author} {\bibfnamefont
  {A.}~\bibnamefont {Sato}}, \bibinfo {author} {\bibfnamefont {K.}~\bibnamefont
  {Yamaura}}, \bibinfo {author} {\bibfnamefont {E.}~\bibnamefont
  {Takayama-Muromachi}}, \bibinfo {author} {\bibfnamefont {H.~F.}\ \bibnamefont
  {Tian}}, \bibinfo {author} {\bibfnamefont {H.~X.}\ \bibnamefont {Yang}},
  \bibinfo {author} {\bibfnamefont {J.~Q.}\ \bibnamefont {Li}}, \bibinfo
  {author} {\bibfnamefont {T.}~\bibnamefont {Varga}}, \bibinfo {author}
  {\bibfnamefont {J.~F.}\ \bibnamefont {Mitchell}}, \ and\ \bibinfo {author}
  {\bibfnamefont {S.}~\bibnamefont {Okamoto}},\ }\href {\doibase
  10.1103/PhysRevB.80.161104} {\bibfield  {journal} {\bibinfo  {journal} {Phys.
  Rev. B}\ }\textbf {\bibinfo {volume} {80}},\ \bibinfo {pages} {161104}
  (\bibinfo {year} {2009})}\BibitemShut {NoStop}%
\bibitem [{\citenamefont {Calder}\ \emph {et~al.}(2012)\citenamefont {Calder},
  \citenamefont {Garlea}, \citenamefont {{McMorrow}}, \citenamefont {Lumsden},
  \citenamefont {Stone}, \citenamefont {Lang}, \citenamefont {Kim},
  \citenamefont {Schlueter}, \citenamefont {Shi}, \citenamefont {Yamaura},
  \citenamefont {Sun}, \citenamefont {Tsujimoto},\ and\ \citenamefont
  {Christianson}}]{calder}%
  \BibitemOpen
  \bibfield  {author} {\bibinfo {author} {\bibfnamefont {S.}~\bibnamefont
  {Calder}}, \bibinfo {author} {\bibfnamefont {V.~O.}\ \bibnamefont {Garlea}},
  \bibinfo {author} {\bibfnamefont {D.~F.}\ \bibnamefont {{McMorrow}}},
  \bibinfo {author} {\bibfnamefont {M.~D.}\ \bibnamefont {Lumsden}}, \bibinfo
  {author} {\bibfnamefont {M.~B.}\ \bibnamefont {Stone}}, \bibinfo {author}
  {\bibfnamefont {J.~C.}\ \bibnamefont {Lang}}, \bibinfo {author}
  {\bibfnamefont {J.~W.}\ \bibnamefont {Kim}}, \bibinfo {author} {\bibfnamefont
  {J.~A.}\ \bibnamefont {Schlueter}}, \bibinfo {author} {\bibfnamefont {Y.~G.}\
  \bibnamefont {Shi}}, \bibinfo {author} {\bibfnamefont {K.}~\bibnamefont
  {Yamaura}}, \bibinfo {author} {\bibfnamefont {Y.~S.}\ \bibnamefont {Sun}},
  \bibinfo {author} {\bibfnamefont {Y.}~\bibnamefont {Tsujimoto}}, \ and\
  \bibinfo {author} {\bibfnamefont {A.~D.}\ \bibnamefont {Christianson}},\
  }\href@noop {} {\bibfield  {journal} {\bibinfo  {journal} {Phys. Rev. Lett.}\
  }\textbf {\bibinfo {volume} {108}},\ \bibinfo {pages} {257209} (\bibinfo
  {year} {2012})}\BibitemShut {NoStop}%
\bibitem [{\citenamefont {Singh}\ and\ \citenamefont
  {Nordstrom}(2006)}]{singh-book}%
  \BibitemOpen
  \bibfield  {author} {\bibinfo {author} {\bibfnamefont {D.~J.}\ \bibnamefont
  {Singh}}\ and\ \bibinfo {author} {\bibfnamefont {L.}~\bibnamefont
  {Nordstrom}},\ }\href@noop {} {\emph {\bibinfo {title} {{Planewaves
  Pseudopotentials and the LAPW Method, 2nd Edition}}}}\ (\bibinfo  {publisher}
  {Springer, Berlin},\ \bibinfo {year} {2006})\BibitemShut {NoStop}%
\bibitem [{\citenamefont {Blaha}\ \emph {et~al.}(2001)\citenamefont {Blaha},
  \citenamefont {Schwarz}, \citenamefont {Madsen}, \citenamefont {Kvasnicka},\
  and\ \citenamefont {Luitz}}]{wien2k}%
  \BibitemOpen
  \bibfield  {author} {\bibinfo {author} {\bibfnamefont {P.}~\bibnamefont
  {Blaha}}, \bibinfo {author} {\bibfnamefont {K.}~\bibnamefont {Schwarz}},
  \bibinfo {author} {\bibfnamefont {G.}~\bibnamefont {Madsen}}, \bibinfo
  {author} {\bibfnamefont {D.}~\bibnamefont {Kvasnicka}}, \ and\ \bibinfo
  {author} {\bibfnamefont {J.}~\bibnamefont {Luitz}},\ }\href@noop {} {\emph
  {\bibinfo {title} {WIEN2k, An Augmented Plane Wave + Local Orbitals Program
  for Calculating Crystal Properties}}}\ (\bibinfo  {publisher} {K. Schwarz,
  Tech. Univ. Wien, Austria},\ \bibinfo {year} {2001})\BibitemShut {NoStop}%
\bibitem [{\citenamefont {Singh}(1991)}]{singh-lo}%
  \BibitemOpen
  \bibfield  {author} {\bibinfo {author} {\bibfnamefont {D.}~\bibnamefont
  {Singh}},\ }\href@noop {} {\bibfield  {journal} {\bibinfo  {journal} {Phys.
  Rev. B}\ }\textbf {\bibinfo {volume} {43}},\ \bibinfo {pages} {6388}
  (\bibinfo {year} {1991})}\BibitemShut {NoStop}%
\bibitem [{\citenamefont {Perdew}\ \emph {et~al.}(1996)\citenamefont {Perdew},
  \citenamefont {Burke},\ and\ \citenamefont {Ernzerhof}}]{pbe}%
  \BibitemOpen
  \bibfield  {author} {\bibinfo {author} {\bibfnamefont {J.~P.}\ \bibnamefont
  {Perdew}}, \bibinfo {author} {\bibfnamefont {K.}~\bibnamefont {Burke}}, \
  and\ \bibinfo {author} {\bibfnamefont {M.}~\bibnamefont {Ernzerhof}},\
  }\href@noop {} {\bibfield  {journal} {\bibinfo  {journal} {Phys. Rev. Lett.}\
  }\textbf {\bibinfo {volume} {77}},\ \bibinfo {pages} {3865} (\bibinfo {year}
  {1996})}\BibitemShut {NoStop}%
\bibitem [{\citenamefont {Mazin}\ and\ \citenamefont {Singh}(1997)}]{mazin-ru}%
  \BibitemOpen
  \bibfield  {author} {\bibinfo {author} {\bibfnamefont {I.~I.}\ \bibnamefont
  {Mazin}}\ and\ \bibinfo {author} {\bibfnamefont {D.~J.}\ \bibnamefont
  {Singh}},\ }\href@noop {} {\bibfield  {journal} {\bibinfo  {journal} {Phys.
  Rev. B}\ }\textbf {\bibinfo {volume} {56}},\ \bibinfo {pages} {2556}
  (\bibinfo {year} {1997})}\BibitemShut {NoStop}%
\bibitem [{\citenamefont {Singh}(1996)}]{singh-ru}%
  \BibitemOpen
  \bibfield  {author} {\bibinfo {author} {\bibfnamefont {D.~J.}\ \bibnamefont
  {Singh}},\ }\href@noop {} {\bibfield  {journal} {\bibinfo  {journal} {J.
  Appl. Phys.}\ }\textbf {\bibinfo {volume} {79}},\ \bibinfo {pages} {4818}
  (\bibinfo {year} {1996})}\BibitemShut {NoStop}%
\bibitem [{\citenamefont {Kosterlitz}\ and\ \citenamefont
  {Thouless}(1973)}]{kt}%
  \BibitemOpen
  \bibfield  {author} {\bibinfo {author} {\bibfnamefont {J.~M.}\ \bibnamefont
  {Kosterlitz}}\ and\ \bibinfo {author} {\bibfnamefont {D.~J.}\ \bibnamefont
  {Thouless}},\ }\href@noop {} {\bibfield  {journal} {\bibinfo  {journal} {J.
  Phys. C}\ }\textbf {\bibinfo {volume} {6}},\ \bibinfo {pages} {1181}
  (\bibinfo {year} {1973})}\BibitemShut {NoStop}%
\bibitem [{\citenamefont {Taylor}\ \emph {et~al.}(2015)\citenamefont {Taylor},
  \citenamefont {Morrow}, \citenamefont {Singh}, \citenamefont {Calder},
  \citenamefont {Lumsden}, \citenamefont {Woodward},\ and\ \citenamefont
  {Christianson}}]{taylor}%
  \BibitemOpen
  \bibfield  {author} {\bibinfo {author} {\bibfnamefont {A.~E.}\ \bibnamefont
  {Taylor}}, \bibinfo {author} {\bibfnamefont {R.}~\bibnamefont {Morrow}},
  \bibinfo {author} {\bibfnamefont {D.~J.}\ \bibnamefont {Singh}}, \bibinfo
  {author} {\bibfnamefont {S.}~\bibnamefont {Calder}}, \bibinfo {author}
  {\bibfnamefont {M.~D.}\ \bibnamefont {Lumsden}}, \bibinfo {author}
  {\bibfnamefont {P.~M.}\ \bibnamefont {Woodward}}, \ and\ \bibinfo {author}
  {\bibfnamefont {A.~D.}\ \bibnamefont {Christianson}},\ }\href@noop {}
  {\bibfield  {journal} {\bibinfo  {journal} {Phys. Rev. B}\ }\textbf {\bibinfo
  {volume} {91}},\ \bibinfo {pages} {100406} (\bibinfo {year}
  {2015})}\BibitemShut {NoStop}%
\bibitem [{\citenamefont {Jacobson}\ \emph {et~al.}(1973)\citenamefont
  {Jacobson}, \citenamefont {Tofield},\ and\ \citenamefont
  {Fender}}]{jacobson}%
  \BibitemOpen
  \bibfield  {author} {\bibinfo {author} {\bibfnamefont {A.~J.}\ \bibnamefont
  {Jacobson}}, \bibinfo {author} {\bibfnamefont {B.~C.}\ \bibnamefont
  {Tofield}}, \ and\ \bibinfo {author} {\bibfnamefont {B.~E.~F.}\ \bibnamefont
  {Fender}},\ }\href@noop {} {\bibfield  {journal} {\bibinfo  {journal} {J.
  Phys. C}\ }\textbf {\bibinfo {volume} {6}},\ \bibinfo {pages} {1615}
  (\bibinfo {year} {1973})}\BibitemShut {NoStop}%
\bibitem [{\citenamefont {Tofield}(1976)}]{tofield}%
  \BibitemOpen
  \bibfield  {author} {\bibinfo {author} {\bibfnamefont {B.}~\bibnamefont
  {Tofield}},\ }\href@noop {} {\bibfield  {journal} {\bibinfo  {journal} {J.
  Phys. Colloq.}\ }\textbf {\bibinfo {volume} {37C6}},\ \bibinfo {pages} {539}
  (\bibinfo {year} {1976})}\BibitemShut {NoStop}%
\bibitem [{\citenamefont {Longo}\ \emph {et~al.}(1968)\citenamefont {Longo},
  \citenamefont {Raccah},\ and\ \citenamefont {Goodenough}}]{longo}%
  \BibitemOpen
  \bibfield  {author} {\bibinfo {author} {\bibfnamefont {J.~M.}\ \bibnamefont
  {Longo}}, \bibinfo {author} {\bibfnamefont {P.~M.}\ \bibnamefont {Raccah}}, \
  and\ \bibinfo {author} {\bibfnamefont {J.~B.}\ \bibnamefont {Goodenough}},\
  }\href@noop {} {\bibfield  {journal} {\bibinfo  {journal} {J. Appl. Phys.}\
  }\textbf {\bibinfo {volume} {39}},\ \bibinfo {pages} {1327} (\bibinfo {year}
  {1968})}\BibitemShut {NoStop}%
\bibitem [{\citenamefont {Bouchard}\ and\ \citenamefont
  {Gillson}(1972)}]{bouchard}%
  \BibitemOpen
  \bibfield  {author} {\bibinfo {author} {\bibfnamefont {R.~J.}\ \bibnamefont
  {Bouchard}}\ and\ \bibinfo {author} {\bibfnamefont {J.~L.}\ \bibnamefont
  {Gillson}},\ }\href@noop {} {\bibfield  {journal} {\bibinfo  {journal}
  {Mater. Res. Bull.}\ }\textbf {\bibinfo {volume} {7}},\ \bibinfo {pages}
  {873} (\bibinfo {year} {1972})}\BibitemShut {NoStop}%
\bibitem [{\citenamefont {Anderson}(1950)}]{anderson}%
  \BibitemOpen
  \bibfield  {author} {\bibinfo {author} {\bibfnamefont {P.~W.}\ \bibnamefont
  {Anderson}},\ }\href@noop {} {\bibfield  {journal} {\bibinfo  {journal}
  {Phys. Rev.}\ }\textbf {\bibinfo {volume} {79}},\ \bibinfo {pages} {350}
  (\bibinfo {year} {1950})}\BibitemShut {NoStop}%
\bibitem [{\citenamefont {Goodenough}(1963)}]{goodenough}%
  \BibitemOpen
  \bibfield  {author} {\bibinfo {author} {\bibfnamefont {J.~B.}\ \bibnamefont
  {Goodenough}},\ }\href@noop {} {\emph {\bibinfo {title} {Magnetism and the
  Chemical Bond}}}\ (\bibinfo  {publisher} {Wiley, New York},\ \bibinfo {year}
  {1963})\BibitemShut {NoStop}%
\bibitem [{\citenamefont {Rhodes}\ and\ \citenamefont
  {Wohlfarth}(1963)}]{rhodes}%
  \BibitemOpen
  \bibfield  {author} {\bibinfo {author} {\bibfnamefont {P.}~\bibnamefont
  {Rhodes}}\ and\ \bibinfo {author} {\bibfnamefont {E.~P.}\ \bibnamefont
  {Wohlfarth}},\ }\href@noop {} {\bibfield  {journal} {\bibinfo  {journal}
  {Proc. Roy. Soc. Lond. A}\ }\textbf {\bibinfo {volume} {273}},\ \bibinfo
  {pages} {247} (\bibinfo {year} {1963})}\BibitemShut {NoStop}%
\bibitem [{\citenamefont {Gunnarsson}(1976)}]{gunnarsson}%
  \BibitemOpen
  \bibfield  {author} {\bibinfo {author} {\bibfnamefont {O.}~\bibnamefont
  {Gunnarsson}},\ }\href@noop {} {\bibfield  {journal} {\bibinfo  {journal} {J.
  Phys. F}\ }\textbf {\bibinfo {volume} {6}},\ \bibinfo {pages} {587} (\bibinfo
  {year} {1976})}\BibitemShut {NoStop}%
\bibitem [{\citenamefont {Mravlje}\ \emph {et~al.}(2012)\citenamefont
  {Mravlje}, \citenamefont {Aichhorn},\ and\ \citenamefont
  {Georges}}]{mravlje}%
  \BibitemOpen
  \bibfield  {author} {\bibinfo {author} {\bibfnamefont {J.}~\bibnamefont
  {Mravlje}}, \bibinfo {author} {\bibfnamefont {M.}~\bibnamefont {Aichhorn}}, \
  and\ \bibinfo {author} {\bibfnamefont {A.}~\bibnamefont {Georges}},\
  }\href@noop {} {\bibfield  {journal} {\bibinfo  {journal} {Phys. Rev. Lett.}\
  }\textbf {\bibinfo {volume} {108}},\ \bibinfo {pages} {197202} (\bibinfo
  {year} {2012})}\BibitemShut {NoStop}%
\end{thebibliography}%

\end{document}